\documentclass[sigconf,natbib=true]{acmart}

\usepackage{booktabs}       
\usepackage{multirow}
\usepackage{graphicx}
\usepackage{amsmath}
\usepackage{bm}             
\usepackage{microtype}      
\usepackage{xcolor}
\usepackage{eurosym}        
\usepackage{tikz}
\usetikzlibrary{arrows.meta,calc,positioning}

\acmConference[CIKM '26]{ACM International Conference on Information
  and Knowledge Management}{2026}{Rome, Italy}
\acmDOI{}
\acmISBN{}
\setcopyright{none}
\copyrightyear{2026}

\title{Probabilistic Salary Prediction with Graph Attention Networks and a Mixture Density Network}

\author{Zhipei Qin}
\affiliation{%
  \institution{Leiden Institute of Advanced Computer Science (LIACS)}
  \institution{Leiden University}
  \city{Leiden}
  \country{The Netherlands}
}
\email{Zhipeiqin17@gmail.com}

\author{Mohammad Shokri}
\affiliation{%
  \institution{Randstad N.V.}
  \city{Amsterdam}
  \country{The Netherlands}
}
\email{mohammad.shokri.acct@gmail.com}

\author{N. van Weeren}
\affiliation{%
  \institution{Leiden Institute of Advanced Computer Science (LIACS)}
  \institution{Leiden University}
  \city{Leiden}
  \country{The Netherlands}
}
\email{n.van.weeren@liacs.leidenuniv.nl}

\author{F.W. Takes}
\affiliation{%
  \institution{Leiden Institute of Advanced Computer Science (LIACS)}
  \institution{Leiden University}
  \city{Leiden}
  \country{The Netherlands}
}
\email{f.w.takes@liacs.leidenuniv.nl}


\begin{abstract}
Accurate salary prediction is critical for bridging the information gap between
employers and job seekers in modern labor markets.
Existing approaches predominantly yield a single point estimate and treat job
attributes such as location, occupation, and industry as independent categorical
features, ignoring both the inherent uncertainty and multi-modality of real-world
compensation data and the rich hierarchical and semantic-similarity relationships
that govern pay norms.
In this paper we propose \textbf{GAT-MDN}, a unified framework that
addresses both limitations simultaneously.
For each of the three attribute domains we construct a domain-specific
graph whose edges encode (i)~\emph{hierarchical} parent--child
containment and (ii)~\emph{weighted similarity} links derived from a
pre-trained Sentence-Transformer.
Parallel Graph Attention Networks (GATs) with edge-feature-aware
attention learn rich, context-sensitive node representations from these
multi-relational graphs.
A \emph{priority-based hierarchical selection} module then assembles a
composite feature vector that gracefully handles missing or coarse
attributes, and a Mixture Density Network (MDN) head maps this vector
to the parameters of a Gaussian Mixture Model (GMM), yielding a full
conditional salary distribution.
Extensive experiments on a real-world Dutch job-posting dataset of over
1\,million records demonstrate that GAT-MDN significantly outperforms
a non-graph MLP-MDN baseline in both Negative Log-Likelihood (NLL) and
Mean Squared Error (MSE).
\end{abstract}


\begin{document}
\maketitle

\section{Introduction}
\label{sec:intro}

\begin{figure*}[t]
\centering
\begin{tikzpicture}[
  font=\small,
  A/.style={-{Stealth[length=4pt,width=3pt]}, semithick},
  BOX/.style={draw, rounded corners=3pt, semithick, align=center, inner sep=4pt},
]

\node[BOX, fill=cyan!12, minimum width=1.8cm, minimum height=0.70cm]
  (inp) at (0,0) {Job Posting\\{\tiny\color{gray}loc / occ / ind}};

\node[BOX, fill=orange!20, minimum width=1.8cm]
  (Gl) at (2.9, 1.2)
  {$G_{\mathrm{loc}}$\\[-2pt]{\tiny\color{gray!60}hier+sim}};
\node[BOX, fill=orange!20, minimum width=1.8cm]
  (Go) at (2.9, 0.0)
  {$G_{\mathrm{occ}}$\\[-2pt]{\tiny\color{gray!60}hier+sim}};
\node[BOX, fill=orange!20, minimum width=1.8cm]
  (Gi) at (2.9,-1.2)
  {$G_{\mathrm{ind}}$\\[-2pt]{\tiny\color{gray!60}hier+sim}};
\node[font=\tiny\itshape, gray, above=5pt of Gl] {Domain Graphs};
\draw[A] (inp.east) to[out=20,in=160]  (Gl.west);
\draw[A] (inp.east) --                  (Go.west);
\draw[A] (inp.east) to[out=-20,in=200] (Gi.west);

\node[BOX, fill=green!18, minimum width=0.85cm, minimum height=0.52cm]
  (G1) at (4.75, 1.2) {GAT};
\node[BOX, fill=green!18, minimum width=0.85cm, minimum height=0.52cm]
  (G2) at (4.75, 0.0) {GAT};
\node[BOX, fill=green!18, minimum width=0.85cm, minimum height=0.52cm]
  (G3) at (4.75,-1.2) {GAT};
\node[font=\tiny\itshape, gray, above=5pt of G1] {Edge-aware GATs};
\draw[A] (Gl)--(G1);
\draw[A] (Go)--(G2);
\draw[A] (Gi)--(G3);

\node[BOX, fill=violet!12, minimum width=1.6cm, minimum height=2.9cm]
  (Sel) at (6.8, 0)
  {Priority\\[-2pt]Selection\\{\tiny\color{gray}finest-first}\\[2pt]{\tiny\color{gray}embed.\ concat}};
\node[font=\tiny\itshape, gray, above=5pt of Sel.north] {Hierarchical};
\draw[A] (G1.east) -- (Sel.west |- G1);
\draw[A] (G2.east) -- (Sel.west);
\draw[A] (G3.east) -- (Sel.west |- G3);

\node[BOX, fill=red!8, minimum width=1.3cm, minimum height=0.65cm]
  (MDN) at (8.7, 0) {MDN\\[-1pt]Head};
\draw[A] (Sel.east)--(MDN.west);

\coordinate (gO) at (9.9, 0.6);
\draw[A,thin] ($(gO)+(0,-0.12)$) -- ($(gO)+(0,1.05)$);
\node[left, font=\tiny] at ($(gO)+(0,0.48)$) {$p(y)$};
\draw[A,thin] ($(gO)+(-0.08,0)$) -- ($(gO)+(2.70,0)$)
  node[right,font=\tiny]{salary};
\draw[very thick, blue!65]
  ($(gO)+(0.00,0.02)$)
  ..controls($(gO)+(0.25,0.02)$)and($(gO)+(0.45,0.88)$)..($(gO)+(0.70,0.88)$)
  ..controls($(gO)+(0.97,0.88)$)and($(gO)+(1.08,0.02)$)..($(gO)+(1.26,0.02)$)
  ..controls($(gO)+(1.44,0.02)$)and($(gO)+(1.58,0.65)$)..($(gO)+(1.80,0.65)$)
  ..controls($(gO)+(2.02,0.65)$)and($(gO)+(2.20,0.02)$)..($(gO)+(2.56,0.02)$);
\node[blue!72, font=\tiny\bfseries] at ($(gO)+(1.35,1.18)$)
  {GAT-MDN: full distribution};
\draw[A] (MDN.east) to[out=15,in=195] ($(gO)+(-0.08,0)$);

\node[font=\normalsize\itshape, gray] at ($(gO)+(1.35,-0.65)$) {vs.};
\node[font=\tiny\itshape, gray!80] at ($(gO)+(1.35,-0.95)$) {existing models:};

\coordinate (pO) at (9.9, -1.90);
\draw[A,thin] ($(pO)+(0,-0.12)$) -- ($(pO)+(0,0.90)$);
\node[left, font=\tiny] at ($(pO)+(0,0.42)$) {$p(y)$};
\draw[A,thin] ($(pO)+(-0.08,0)$) -- ($(pO)+(1.25,0)$);
\draw[very thick, red!65]
  ($(pO)+(0.60,0.00)$) -- ($(pO)+(0.60,0.76)$);
\fill[red!65] ($(pO)+(0.60,0.76)$) circle (2.2pt);
\node[red!70, font=\tiny\bfseries] at ($(pO)+(0.60,1.05)$)
  {Point estimate $\hat{y}$};

\coordinate (uO) at (11.45, -1.90);
\draw[A,thin] ($(uO)+(0,-0.12)$) -- ($(uO)+(0,0.90)$);
\node[left, font=\tiny] at ($(uO)+(0,0.42)$) {$p(y)$};
\draw[A,thin] ($(uO)+(-0.08,0)$) -- ($(uO)+(1.55,0)$)
  node[right,font=\tiny]{salary};
\draw[very thick, orange!80!black]
  ($(uO)+(0.00,0.02)$)
  ..controls($(uO)+(0.22,0.02)$)and($(uO)+(0.52,0.82)$)..($(uO)+(0.78,0.82)$)
  ..controls($(uO)+(1.04,0.82)$)and($(uO)+(1.36,0.02)$)..($(uO)+(1.50,0.02)$);
\node[orange!85!black, font=\tiny\bfseries] at ($(uO)+(0.75,1.05)$)
  {Unimodal dist.};

\end{tikzpicture}
\caption{Overview of GAT-MDN and the core problem it addresses.
  \emph{Pipeline (left to right):} a job posting with three attribute
  domains is processed through domain-specific graphs encoding hierarchical
  parent--child relations and weighted semantic-similarity links; parallel
  edge-aware Graph Attention Networks produce context-sensitive node
  embeddings; a priority-based hierarchical selection module concatenates
  the embeddings from the finest available granularity; an MDN head maps
  this to a Gaussian Mixture Model.
  \emph{Key contrast (right):} GAT-MDN outputs a full conditional salary
  distribution capturing multiple salary modes (blue), whereas existing
  models reduce to either a single point estimate~$\hat{y}$ (red) or a
  unimodal distribution (orange) that cannot capture multi-modal pay norms.}
\label{fig:overview}
\end{figure*}
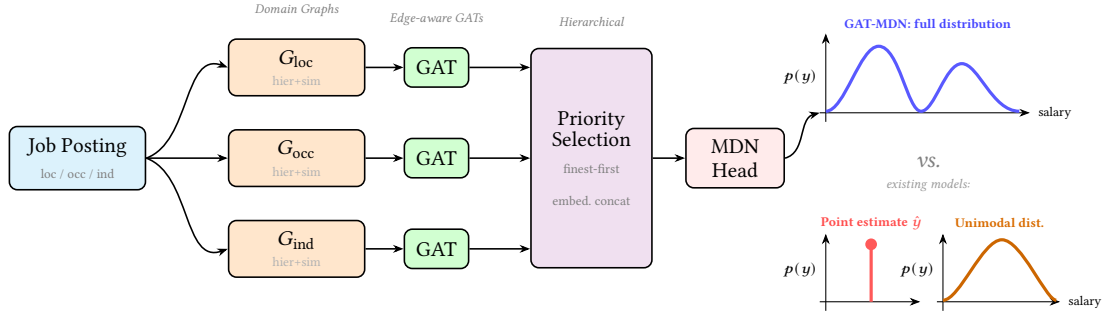

The Human Resources (HR) and recruitment industry plays an indispensable
intermediary role in the labor market, matching talent to organizational
needs with salary as the primary currency of exchange.
A competitive and transparent compensation offer allows employers to
attract the right talent while controlling operational costs; for
candidates, it enables informed career decisions and fair wage
negotiation.
Yet a persistent \emph{salary information gap}~\cite{meng2018salary}
means that applicants rarely know the market rate for a position until
an offer arrives, while employers risk either losing candidates with
uncompetitive bids or overpaying relative to market norms.
Bridging this gap requires models that can accurately estimate not just
an expected salary, but the \emph{full distribution} of plausible
compensation for any given combination of job attributes.

First, the majority of data-driven salary prediction methods, including
linear regression~\cite{bhat2025community}, tree
ensembles~\cite{zhalilova2024forecasting}, and deep text
models~\cite{wang2019bi}, produce a single scalar forecast.
In practice, salaries exhibit a \emph{multi-modal} and \emph{skewed}
distribution: the same job title can attract very different compensation
depending on seniority, sector, and local economic conditions.
A single point estimate conceals this complexity, conveys no
uncertainty, and is ill-suited for downstream decision support tasks
such as budget planning, offer benchmarking, or risk assessment.

Second, standard machine learning pipelines encode job attributes such as location,
occupation, and industry as flat, independent categorical features,
typically via one-hot encoding or label encoding.
The location domain is explicitly hierarchical: provinces contain
cities, and salary norms propagate along this containment tree.
The occupation domain follows a three-level hierarchy and harbors rich
semantic similarities: a ``Data Scientist'' and a ``Data Analyst''
share skills and market value in ways that raw categorical codes cannot
capture.
The industry domain is similarly structured under the international
NACE Rev.~2 standard~\cite{nace2008}.
Treating these inter-related attributes as independent features discards
context that is directly predictive of salary, leading to systematic
under-fitting for rare or unseen attribute combinations.

Third, real-world job postings are frequently incomplete.
An employer may specify only a province rather than a city, or list a
broad occupational category rather than a precise title.
A production-grade salary estimator must degrade gracefully when the
finest-grained attribute is unavailable, falling back to progressively
coarser representations without losing predictive power.

To address all three challenges jointly, we propose \textbf{GAT-MDN},
a framework that combines multi-relational Graph Attention Networks with
a Mixture Density Network (Figure~\ref{fig:overview}).
For each of the three attribute domains (Location, Occupation, Industry),
we construct a domain-specific graph whose edges encode both hierarchical
parent--child containment and weighted semantic similarity between
peer-level nodes derived from a pre-trained
Sentence-Transformer~\cite{reimers2019sentencebert}.
Parallel Graph Attention Networks~\cite{velickovic2018gat} with an
edge-feature-aware attention mechanism learn context-sensitive node
representations.
A priority-based hierarchical selection module assembles a composite
feature vector by retrieving the finest-grained available embedding for
each domain, and a Mixture Density Network head maps this vector to the
parameters of a multi-component Gaussian Mixture Model, directly
outputting the full conditional salary distribution.

In summary, this paper makes the following contributions:
\begin{enumerate}
\item \textbf{Dual-edge multi-graph construction.}
We introduce a principled graph construction strategy that encodes
\emph{both} hierarchical containment and semantic similarity as
distinct, edge-typed relations, providing the downstream GNN with
a richer structural context than prior single-relation graph
approaches~\cite{chen2020gcnsalary}.

\item \textbf{Edge-aware GAT for attribute representation learning.}
We extend the standard GAT~\cite{velickovic2018gat} with an
edge-feature attention term and apply it in parallel across three
domain-specific graphs, learning node embeddings that capture
market-level context beyond what flat categorical features convey.

\item \textbf{Priority-based hierarchical selection.}
We design a coarse-to-fine feature selection mechanism that enables the
model to handle inputs at any granularity level without requiring
complete attribute specifications, improving real-world applicability.

\item \textbf{Probabilistic salary distribution prediction.}
By coupling the GAT feature extractor with an MDN head, our framework
outputs a full Gaussian Mixture Model distribution over salary,
quantifying uncertainty and capturing multi-modality that point-estimate
models miss entirely.

\item \textbf{Large-scale empirical validation.}
We evaluate GAT-MDN on over 1\,million real-world Dutch job postings,
with ablation studies establishing quantitative superiority over a
non-graph baseline.
\end{enumerate}

\section{Related Work}
\label{sec:related}

\subsection{Data-Driven Salary Prediction}
Salary prediction has been studied using diverse methodologies.
Early approaches employ linear regression or traditional ML methods
such as SVM~\cite{svm_income} and decision trees~\cite{zhalilova2024forecasting}.
Deep learning approaches have since improved predictive accuracy:
Wang et al.~\cite{wang2019bi} design a Bi-GRU-CNN architecture for
salary prediction from job description text.
Meng et al.~\cite{meng2018salary} frame salary benchmarking as matrix
completion (HSBMF), while a follow-up~\cite{meng2022ndp} addresses
data sparsity with a Dirichlet-process prior.
Sun et al.~\cite{sscn2021} propose the Salary-Skill Combination Network
to disentangle the value of individual skills, and LGDESetNet~\cite{lgde2022}
identifies influential skill subsets via a graph-enhanced disentangled
selection layer.
Despite their diversity, all of these works share one fundamental
limitation: they output a single point estimate, making them incapable
of expressing the inherent uncertainty and multi-modality of real
salary distributions.

\subsection{Graph-Based Models for Job Market Data}
Graph neural networks have demonstrated strong performance on relational
and heterogeneous data.
HetGNN~\cite{hetgnn2019} handles heterogeneous graphs by sampling
neighbor types and aggregating their features.
Chen et al.~\cite{chen2020gcnsalary} propose a semi-supervised GCN for
salary estimation that leverages both structural and textual features
from job postings, demonstrating the value of graph structure for this
task.
Our model employs Graph Attention Networks~\cite{velickovic2018gat}
for their ability to learn adaptive attention weights, which is
crucial given that our graphs contain edges of two distinct semantic
types (hierarchical and similarity-based).
To our knowledge, ours is the first work to construct multi-relational
graphs explicitly encoding \emph{both} hierarchical containment
\emph{and} semantic similarity between job attributes.

\subsection{Mixture Density Networks and Probabilistic Prediction}
The Mixture Density Network was originally proposed by
Bishop~\cite{bishop1994mdnoriginal} to output the parameters of a
Gaussian Mixture Model from a neural network.
MDNs have since been applied to speech enhancement~\cite{mdn_speech},
handwriting synthesis~\cite{graves2013handwriting}, and motion
prediction.
Gaussian Mixture Models serve dual roles as semi-parametric density
estimators and clustering tools~\cite{gmm_review}.
VaDE~\cite{vade2017} embeds a GMM prior in a Variational Autoencoder
for joint representation learning and clustering.
Our model is most closely related to this deep-GMM paradigm but, to
our knowledge, is the first to pair a multi-graph GAT with an MDN head
for structured salary prediction.

\section{Preliminaries}
\label{sec:prelim}

\subsection{Graph-Based Data Representation}
\label{sec:prelim:graph}

\noindent\textbf{Formal notation.}
Let $\mathcal{A} = \{a_1, a_2, \ldots, a_N\}$ denote the set of
attribute values characterising a single job instance, belonging to
$M{=}3$ domains: Location ($m_\text{loc}$), Occupation ($m_\text{occ}$),
and Industry ($m_\text{ind}$).
For each domain $m$ we construct a graph
$\mathcal{G}_m = (\mathcal{V}_m, \mathcal{E}_m)$,
where $\mathcal{V}_m$ is the set of unique attribute values (nodes) and
$\mathcal{E}_m = \mathcal{E}^\text{hier}_m \cup \mathcal{E}^\text{sim}_m$
is the union of hierarchical and similarity edges.

\textbf{Hierarchical edges} $\mathcal{E}^\text{hier}_m$ encode
parent--child co-occurrence across adjacent hierarchy levels:
\begin{equation}
  \mathcal{E}^\text{hier}_m =
  \bigl\{\{u,v\} \;\big|\; u = d_i(L_{m,p}),\; v = d_i(L_{m,c}),\;
         u,v \neq \texttt{null}\bigr\}_{d_i \in \mathcal{D}}
\end{equation}
$L_{m,p}$ and $L_{m,c}$ represent the parent and child levels in the
hierarchy; $d_i$ is a specific job posting in the dataset.

\textbf{Weighted similarity edges} $\mathcal{E}^\text{sim}_m$ connect
peer-level nodes whose Sentence-Transformer embeddings $\phi(\cdot)$
satisfy cosine similarity above threshold $\theta$:
\begin{equation}
  \mathcal{E}^\text{sim}_m =
  \bigl\{(u,v,w_{uv}) \;\big|\;
  u,v \in \mathcal{V}_{m,l},\; u \neq v,\;
  w_{uv} = \mathrm{sim}(\phi(u),\phi(v)) > \theta \bigr\}
\end{equation}
$\mathcal{V}_{m,l}$ is the set of nodes at level $l$ in domain $m$.

\subsection{Probabilistic Prediction Goal}
\label{sec:prelim:goal}
Given attribute set $\mathcal{A}$ and graphs $\{\mathcal{G}_m\}$, the
model learns $f$ that predicts the conditional salary distribution as a
$K$-component GMM:
\begin{equation}
  \hat{p}(y \mid \mathcal{A}) =
  \sum_{k=1}^{K} \pi_k(\mathcal{A};\Theta)\,
  \mathcal{N}\!\left(y \;\big|\; \mu_k(\mathcal{A};\Theta),
                                 \sigma_k^2(\mathcal{A};\Theta)\right)
\end{equation}
where $\pi_k \geq 0$, $\sum_k \pi_k = 1$.

\subsection{Training Objective}
\label{sec:prelim:loss}
The model is trained end-to-end by minimising:
\begin{equation}
  \mathcal{L} = \mathcal{L}_\text{NLL} + \lambda\,\mathcal{L}_\text{MSE}
\end{equation}
where
$\mathcal{L}_\text{NLL} = -\sum_i \log \hat{p}(y^{(i)} \mid \mathcal{A}^{(i)})$
is the negative log-likelihood loss,
$\mathcal{L}_\text{MSE} = \sum_i (y^{(i)} - \mathbb{E}[y|\mathcal{A}^{(i)}])^2$
is an auxiliary MSE term that anchors the distribution mean, and
$\lambda$ is a balancing hyperparameter.

\section{Methodology}
\label{sec:method}

\subsection{Graph Construction}
\label{sec:method:graph}

Raw salary bounds $S$ are first transformed via log-standardisation
$S' = (\log S - \mu_{\log}) / \sigma_{\log}$, yielding a near-normal
training target.
For each domain $m$, unique attribute values across all hierarchical
levels are collected; each is added as a node in $\mathcal{G}_m$ with
its level stored as a node attribute.

\textbf{Hierarchical edge construction.}
For every job record, co-occurring non-null values at adjacent
hierarchical levels generate an undirected edge, capturing formal
containment relations without a hand-crafted ontology.

\textbf{Similarity edge construction.}
Within each level, the \path{all-MiniLM-L6-v2} Sentence-Transformer
encodes each node name into a 384-dimensional vector.
Pairwise cosine similarities are computed and edges are added between
nodes exceeding threshold $\theta = 0.5$, connecting semantically
related roles (e.g., \textit{Data Scientist}~$\leftrightarrow$~\textit{Data Analyst}).

\subsection{GAT Feature Learning}
\label{sec:method:gat}

Each node $v_i$ is assigned an augmented feature vector
$\tilde{x}_i = [x_i \| e_i]$,
where $x_i \in \mathbb{R}^{N_m}$ is a one-hot identity vector and
$e_i \in \mathbb{R}^{d_e}$ is a learnable embedding.
A single GAT layer computes attention weight $\alpha_{ij}$ for each
edge $(i,j) \in \mathcal{E}_m$ as:
\begin{equation}
  \alpha_{ij} = \mathrm{LeakyReLU}
  \!\left(\mathbf{a}^\top
  [W\tilde{x}_i \| W\tilde{x}_j \| W_E x_{ij}]\right)
\end{equation}
where $x_{ij}$ is the edge feature vector encoding the edge type
(hierarchical vs.\ similarity) and edge weight, and $W_E$ is a
learnable edge transformation.
Normalised attention weights $\hat{\alpha}_{ij} = \mathrm{softmax}_j(\alpha_{ij})$
aggregate neighbour features:
$h_i = \sigma\!\left(\sum_{j\in\mathcal{N}_i} \hat{\alpha}_{ij} W\tilde{x}_j\right)$.
Multi-head attention is applied for intermediate layers, with
outputs concatenated and projected to output dimension $F_\text{out}$.
Running a dedicated GAT module on each domain graph $\mathcal{G}_m$
yields a node representation matrix
$H_m \in \mathbb{R}^{N_m \times F_\text{out}}$.

\subsection{Priority-Based Hierarchical Selection}
\label{sec:method:select}
For a given input sample, the final composite feature is assembled by
retrieving one embedding per domain and concatenating:
$\mathbf{h}_\text{comb} = [h_\text{loc} \| h_\text{occ} \| h_\text{ind}]$.
For each domain, the selection follows a \emph{finest-first} priority:
the most granular available attribute value is used (e.g., city-level
takes precedence over province-level in Location); if no attribute is
available, the domain falls back to a learnable ``unknown'' embedding.
This mechanism enables graceful handling of partially specified queries
without any additional imputation step.

\subsection{Mixture Density Network Head}
\label{sec:method:mdn}
The MDN head receives $\mathbf{h}_\text{comb}$ and produces GMM
parameters through a two-layer MLP:
\begin{itemize}
  \item \textbf{Means} $\mu_k$: linear activation.
  \item \textbf{Standard deviations} $\sigma_k$: exponential activation
        ($\sigma_k = \exp(a^{\sigma}_k)$) to guarantee positivity.
  \item \textbf{Mixture weights} $\pi_k$: Softmax over $K$ logits.
\end{itemize}
Together these parameterise the full conditional distribution
$\hat{p}(y \mid \mathcal{A})$ defined in Section~\ref{sec:prelim}.

\section{Descriptive Data Analysis}
\label{sec:data}

Our dataset is sourced from Jobdigger~\cite{jobdigger}, a Dutch
labor-market data platform, and covers January 2022 to January 2025.
After preprocessing, the working corpus comprises \textbf{1,040,918}
salary records with a mean annual salary of \euro{}42,604.
The raw salary distribution is right-skewed and multi-modal; log
transformation produces a near-normal distribution, confirming the
suitability of the chosen preprocessing pipeline.

Table~\ref{tab:graph_stats} summarises the three constructed graphs.

\begin{table}[h]
  \centering
  \caption{Statistics of the three constructed domain graphs.}
  \label{tab:graph_stats}
  \begin{tabular}{lrrr}
    \toprule
    \textbf{Graph} & \textbf{Hierarchy} & \textbf{Nodes} & \textbf{Edges} \\
    \midrule
    Location    & Province / City                    & 478   & 10,177  \\
    Occupation  & Major / Sub / Minor (3 levels)     & 3,128 & 605,767 \\
    Industry    & Major / Sub (NACE Rev.~2)           & 98    & 1,159   \\
    \bottomrule
  \end{tabular}
\end{table}

\noindent
\textbf{Location.}
Among the 12 Dutch provinces, the highest-paying is Zuid-Holland
(\euro{}46K+) and the lowest is Drenthe (\euro{}36K), a 37\% gap.
At the finer city level this spread widens to more than $6\times$,
illustrating the value of fine-grained geographic signals.

\noindent
\textbf{Industry.}
Among the 20 major NACE sectors, the highest-paying is Government
\& Military (\euro{}51,501) and the lowest is Hospitality
(\euro{}31,336), an 82\% gap, with Environmental Services
(\euro{}50,844) and ICT (\euro{}48,620) also among the top sectors.

\noindent
\textbf{Occupation.}
Among the major occupational categories, the highest-paying is Finance
(\euro{}63,105) and the lowest is Cleaning (\euro{}24,461), a gap of
more than 158\%, with ICT (\euro{}52,819) and Social Care
(\euro{}52,138) also near the top.

\section{Experiments}
\label{sec:exp}

We compare GAT-MDN quantitatively against a non-graph ablation
baseline to measure the isolated contribution of the graph structure.

\subsection{Experimental Setup}
\label{sec:exp:setup}
All experiments were run on a single server equipped with one
NVIDIA RTX~4000 Ada GPU, 16 vCPUs, and 62~GB of RAM under
Windows Server 2022, Python~3.10.
The deep learning stack uses PyTorch and PyTorch Geometric;
graph construction relies on NetworkX; semantic embeddings are
generated with \path{sentence-transformers}~\cite{reimers2019sentencebert}.
Hyperparameter search is managed by Optuna~\cite{optuna2019}.
Data is split 80\% train / 20\% validation; both models are trained
for 500 epochs.

\subsection{Hyperparameters}
\label{sec:exp:hyper}
Table~\ref{tab:hyperparams} reports the fixed architectural parameters.
The remaining hyperparameters---learning rate and number of Gaussian
components $K$---are tuned automatically with Optuna using 60~trials
and 100-epoch pruned training.
The search ranges are learning rate $\in \{5\times10^{-4}, 10^{-4},
5\times10^{-5}, 10^{-5}, 5\times10^{-6}\}$ and $K \in [1,10]$.
The optimal configuration for GAT-MDN is $\text{lr}=10^{-4}$, $K=9$;
for the MLP-MDN baseline, $\text{lr}=10^{-4}$, $K=7$.

\begin{table}[h]
  \centering
  \caption{Fixed model hyperparameters.}
  \label{tab:hyperparams}
  \setlength{\tabcolsep}{4pt}
  \begin{tabular}{lr}
    \toprule
    \textbf{Parameter}           & \textbf{Value} \\
    \midrule
    Embedding dim $d_e$          & 32             \\
    GAT hidden channels          & 64             \\
    GAT output dim $F_\text{out}$ & 32            \\
    Attention heads $K_a$         & 3             \\
    MDN hidden dim               & 128            \\
    GAT layers                   & 2              \\
    Batch size                   & 512            \\
    MSE weight $\lambda$         & 0.5            \\
    Optimiser                    & Adam           \\
    \bottomrule
  \end{tabular}
\end{table}

\subsection{Ablation Baseline: MLP-MDN}
\label{sec:exp:baseline}
To isolate the contribution of graph convolution, we implement an
\textbf{Embedding-only MLP-MDN} that replaces the GAT modules with a
standard MLP operating on the raw concatenated node embeddings,
i.e., $\mathbf{h}_\text{comb}^\text{MLP} = [e_\text{loc} \| e_\text{occ} \| e_\text{ind}]$.
This baseline retains the same MDN head, embedding dimensions, and
training procedure, ensuring that any performance gap is attributable
solely to the graph convolution.

\subsection{Salary Prediction Results}
\label{sec:exp:results}
\begin{figure}[t]
  \centering
  \includegraphics[width=0.48\linewidth]{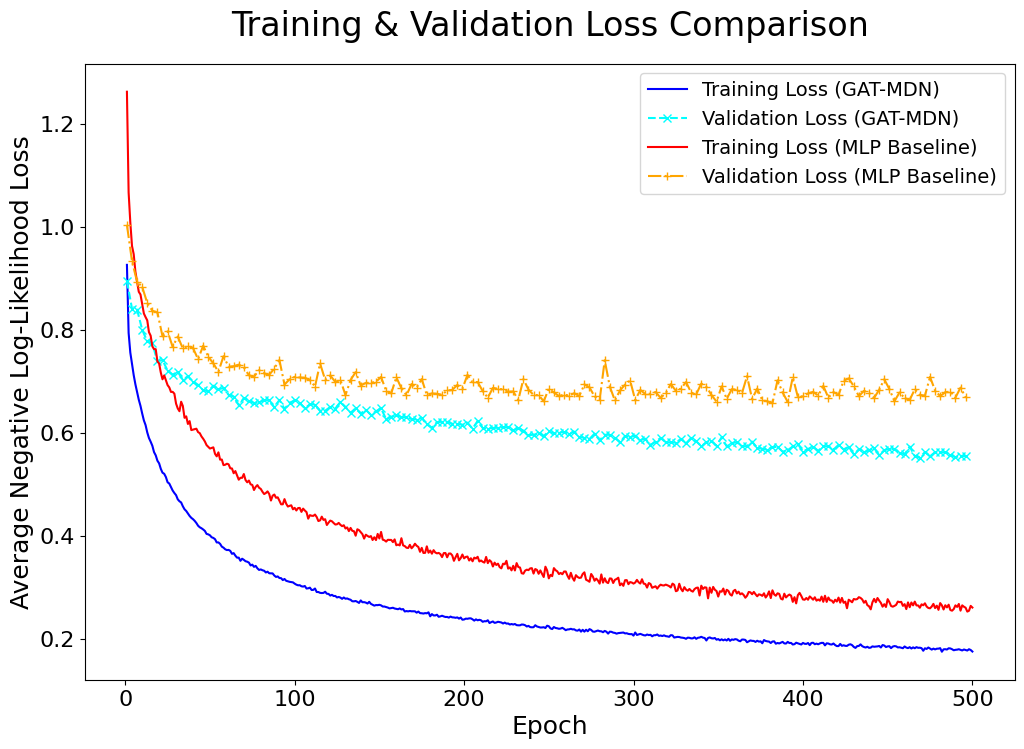}
  \includegraphics[width=0.48\linewidth]{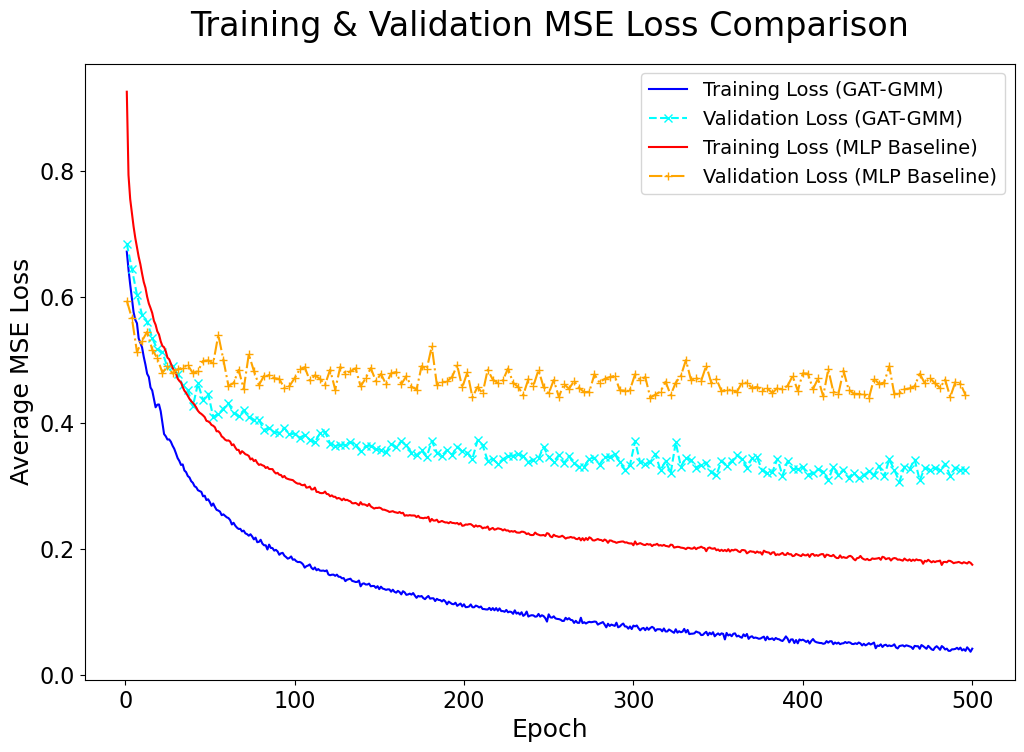}
  \caption{Training and validation NLL (left) and MSE (right) loss curves
    for GAT-MDN vs.\ MLP-MDN over 500 epochs.
    GAT-MDN achieves persistently lower losses on both metrics.}
  \label{fig:loss}
\end{figure}

Figure~\ref{fig:loss} shows the NLL and MSE learning curves for both
models over 500 epochs.
From the earliest epochs, GAT-MDN achieves substantially lower training
NLL.
Beyond the 200-epoch mark, the MLP-MDN baseline stagnates on the
validation NLL while GAT-MDN continues to improve, widening the gap
throughout training.
Analogous superiority is observed in the MSE loss.
These results strongly validate our central hypothesis: explicitly
modelling attribute hierarchy and semantic similarity via graph
attention provides significant predictive benefit over unstructured
embedding learning.

\section{Conclusion}
\label{sec:conclusion}

We presented \textbf{GAT-MDN}, a framework for probabilistic salary
prediction that overcomes two fundamental limitations of prior work:
the reliance on scalar point estimates and the neglect of rich
relational structure among job attributes.
By constructing multi-relational domain graphs with dual (hierarchical
and similarity) edges, training parallel edge-aware GATs for contextual
attribute representation, and coupling the learned representations to
an MDN head, GAT-MDN directly predicts the full conditional salary
distribution.
A priority-based hierarchical selection mechanism ensures graceful
degradation under incomplete inputs.
Experiments on over 1\,million real Dutch job postings confirm
significant improvements over a non-graph ablation baseline in both
NLL and MSE.

Future directions include: (i) incorporating numerical attributes such
as years of experience; (ii) replacing the GMM head with a Normalizing
Flow for more expressive density estimation; (iii) adopting
LLM-initialised node features~\cite{he2023llmgnn} to generalise to
unseen attribute values.

\bibliographystyle{ACM-Reference-Format}
\bibliography{ref}

\end{document}